\documentclass[runningheads]{llncs}
\usepackage[T1]{fontenc}
\usepackage{graphicx,verbatim}
\usepackage{booktabs}
\usepackage{hyperref}
\usepackage{color}

\usepackage{graphicx}
\usepackage{booktabs}
\usepackage{amssymb}
\usepackage{amsmath} 
\begin{document}
\title{Task-oriented Uncertainty Collaborative Learning for Label-Efficient Brain Tumor Segmentation}
\titlerunning{TUCL Framework for Label-Efficient Brain Tumor Segmentation}
%
\begin{comment}  %% Removed for anonymized MICCAI 2025 submission
\author{First Author\inst{1}\orcidID{0000-1111-2222-3333} \and
Second Author\inst{2,3}\orcidID{1111-2222-3333-4444} \and
Third Author\inst{3}\orcidID{2222--3333-4444-5555}}
%
\authorrunning{F. Author et al.}
% First names are abbreviated in the running head.
% If there are more than two authors, 'et al.' is used.
%
\institute{Princeton University, Princeton NJ 08544, USA \and
Springer Heidelberg, Tiergartenstr. 17, 69121 Heidelberg, Germany
\email{lncs@springer.com}\\
\url{http://www.springer.com/gp/computer-science/lncs} \and
ABC Institute, Rupert-Karls-University Heidelberg, Heidelberg, Germany\\
\email{\{abc,lncs\}@uni-heidelberg.de}}

\end{comment}

\author{Zhenxuan Zhang\inst{1}* \and
Hongjie Wu\inst{2}\thanks{Equal contribution.} \and
Jiahao Huang \inst{1, 3} \and
Baihong Xie\inst{6} \and
Zhifan Gao\inst{6} \and
Junxian Du\inst{1} \and
Pete Lally\inst{1} \and
Guang Yang\inst{1,3,4,5}
}
\authorrunning{Zhenxuan Zhang et al.}
\institute{Bioengineering Department and Imperial-X, Imperial College London, London W12 7SL, UK \and
Department of Computing, Imperial College London, London SW7 2AZ, UK\and
National Heart and Lung Institute, Imperial College London, London SW7 2AZ, UK \and
Cardiovascular Research Centre, Royal Brompton Hospital, London SW3 6NP, UK \and
School of Biomedical Engineering \& Imaging Sciences, King's College London, London WC2R 2LS, UK \and
School of Biomedical Engineering, Sun Yat-sen University, Guangzhou 510006, China\\\email{g.yang@imperial.ac.uk}}

\maketitle              % typeset the header of the contribution
\begin{abstract}
Multi-contrast magnetic resonance imaging (MRI) plays a vital role in brain tumor segmentation and diagnosis by leveraging complementary information from different contrasts. Each contrast highlights specific tumor characteristics, enabling a comprehensive understanding of tumor morphology, edema, and pathological heterogeneity. However, existing methods still face the challenges of multi-level specificity perception across different contrasts, especially with limited annotations. These challenges include data heterogeneity, granularity differences, and interference from redundant information. To address these limitations, we propose a Task-oriented Uncertainty Collaborative Learning (TUCL) framework for multi-contrast MRI segmentation. TUCL introduces a task-oriented prompt attention (TPA) module with intra-prompt and cross-prompt attention mechanisms to dynamically model feature interactions across contrasts and tasks. Additionally, a cyclic process is designed to map the prediction back to the prompt to ensure that the prompts are effectively utilized. In the decoding stage, the TUCL framework proposes a dual-path uncertainty refinement (DUR) strategy which ensures robust segmentation by refining predictions iteratively. Extensive experimental results on limited labeled data demonstrate that TUCL significantly improves segmentation accuracy (88.2\% in Dice and 10.853 mm in HD95). It shows that TUCL has the potential to extract multi-contrast information and reduce the reliance on extensive annotations. The code is available at: \url{https://github.com/Zhenxuan-Zhang/TUCL_BrainSeg}.

% Authors must provide keywords and are not allowed to remove this Keyword section.

\end{abstract}
\section{Introduction}

% Multi-contrast collaborative analysis in magnetic resonance imaging (MRI) is crucial for brain tumor segmentation and diagnosis \cite{clinical_1,clinical_2}. In clinical practice, multi-contrast modalities include T1-weighted, T2-weighted, fluid-attenuated inversion recovery (FLAIR), contrast-enhanced T1-weighted imaging (T1ce), etc \cite{clinical_3}. These modalities enable clinicians to achieve a more comprehensive understanding of tumor morphology, edema, and pathological heterogeneity (e.g., T1-weighted images highlight anatomical structures, while FLAIR is particularly effective in delineating peritumoral edema and non-enhancing tumor regions) \cite{brain_fusion,mmformer}. The accurate segmentation of the tumor region can further assist in quantifying tumor volume and edema volume \cite{swinbts}. This is critical for accurate diagnosis, surgical planning, and radiotherapy target delineation. However, manually segmenting brain tumors across multiple contrasts is labor-intensive and subject to inter-observer variability. This limits the application of multi-contrast MRI in tumor segmentation and diagnosis \cite{brain_fusion,brats_2021,swinbts,wenxuan2021transbts}. There is a need to design automated methods that leverage the information provided by multi-contrast MRI for robust and precise tumor segmentation.
Multi-contrast collaborative analysis in magnetic resonance imaging (MRI) is crucial for brain tumor segmentation and diagnosis \cite{clinical_1,clinical_2}. In clinical practice, multi-contrast modalities include T1-weighted, T2-weighted, fluid-attenuated inversion recovery (FLAIR), contrast-enhanced T1-weighted imaging (T1ce), etc \cite{clinical_3}. These modalities allow clinicians to gain a more comprehensive understanding of tumor morphology, edema, and pathological heterogeneity (e.g., FLAIR is particularly effective in delineating peritumoral edema) \cite{brain_fusion,mmformer}. The accurate segmentation of the tumor region can further assist in quantifying tumor volume and edema volume \cite{swinbts}. These aspects are critical for accurate diagnosis, surgical planning, and radiotherapy target delineation. However, manually segmenting brain tumors across multiple contrasts is labor-intensive. This limits the application of multi-contrast MRI in tumor segmentation and diagnosis \cite{brain_fusion,brats_2021,swinbts,wenxuan2021transbts}. Furthermore, obtaining high-quality annotations for all contrasts is costly and time-consuming. As a result, label-efficient segmentation approaches that can effectively utilize limited annotated data while leveraging the complementary information from multi-contrast MRI are highly desirable \cite{nie2019difficulty}. There is a need to design automated methods that leverage the information provided by multi-contrast MRI for robust and precise tumor segmentation while reducing annotation burdens.

\begin{figure*}[t]
\centerline{\includegraphics[width=\columnwidth]{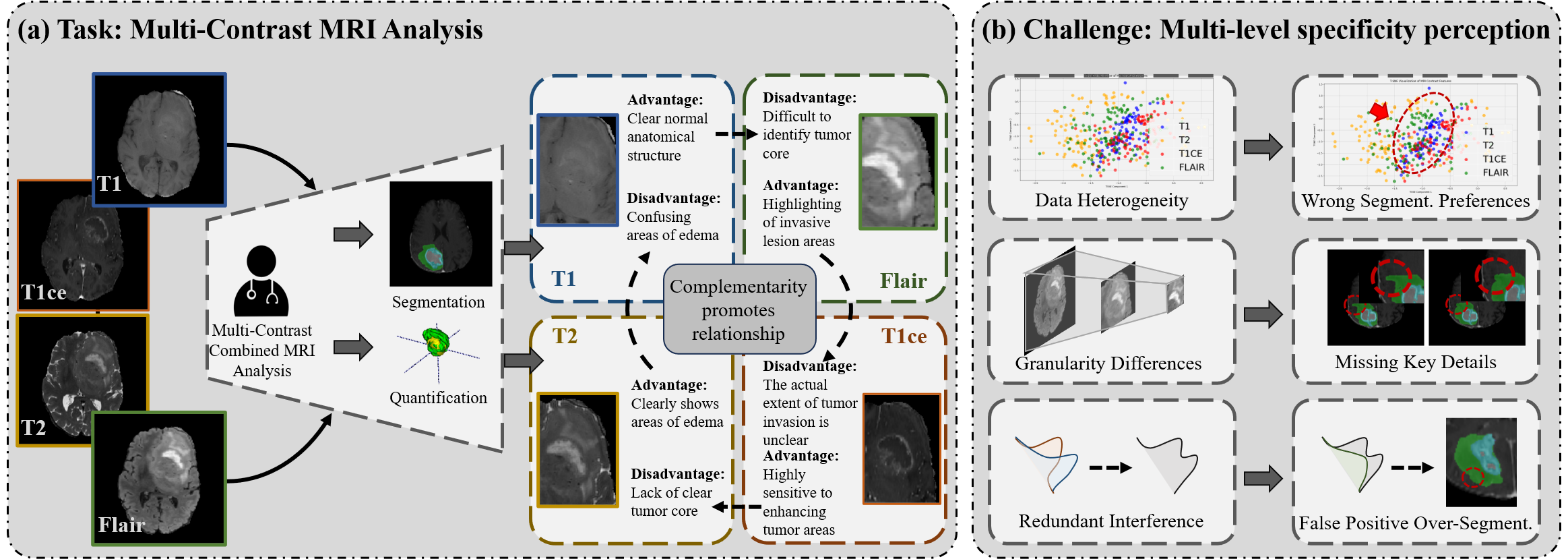}}
\caption{Motivation of our TUCL framework. (a) Task: Multi-Contrast MRI Analysis. It aims to leverage complementary information from T1, T1ce, T2, and FLAIR contrasts for segmentation and further quantification. (b) Challenge: Multi-Level Specificity. It aims to address the challege of data heterogeneity, granularity differences, and redundant interference to improve analysis accuracy.}
\label{fig1}
\end{figure*}

However, collaborative analysis of multi-contrast MRI data still faces significant challenges related to multi-level specificity perception \cite{multi-modal-1,multi-modal-2,multi-modal-3,brain_fusion}.  These challenges primarily stem from three aspects: data heterogeneity, granularity differences, and redundant information interference. Firstly, data heterogeneity arises due to the inherent differences between various MRI contrasts. These modalities capture unique aspects of tumor features (e.g., T1 weighted contrast provides clear anatomical structures) \cite{brain_fusion,brats_2021}. These contrasts exhibit different sensitivities to specific tissue types, making it challenging to integrate different data sources into a unified representation without losing clinical information. Secondly, granularity differences come from different reflections on tumor details. It may affect the fusion of fine-textured microstructures to broader-scale anatomical features (e.g., the fine structural details of a tumor core may be more clearly visible in T1ce images). It requires the appropriate perception of these differences without distorting key tumor information. Third, redundant information interference presents a significant barrier in multi-contrast MRI analysis (e.g., T1- and T2-weighted images might highlight similar tumor boundaries). It leads to potential redundancies in the data. These redundancies may lead to unnecessary or conflicting features that may affect the accuracy of tumor segmentation. 

Existing methods struggle to effectively address these challenges. Traditional methods (e.g., simple concatenation or early fusion) fail to capture the complex interactions among different contrasts \cite{zhang2023deep}. These methods treat each contrast as an independent source of information, which overlooks the contextual relationships between them \cite{wang2022multimodal}. This results in an incomplete representation of the tumor. Recent methods like multi-contrast cross attention mechanisms still have limitations in balancing the contributions of each image contrast \cite{10377125,huang2023accurate}. These methods often focus on global feature alignment but may miss the importance of task-specific refinements (e.g. specific tumor region details might be more clearly visible in one contrast but harder to detect in others) \cite{zhang2022mmformer}. In these cases, they fail to preserve the nuanced and localized details that are essential for accurate segmentation \cite{dolz2018hyperdense}. This can lead to inaccurate tumor boundaries or incorrect identification of regions of necrosis. Lastly, existing methods may not handle the redundancy interaction of multi-contrast data due to the lack of constraints on redundant boundaries \cite{rahman2021redundancy,de2024modality}. The redundant features can be amplified and cause conflicting features (i.e. similar tumor boundaries in two contrasts are interpreted differently and cause inconsistent segmentation).

\begin{figure*}[t]
\centerline{\includegraphics[width=\columnwidth]{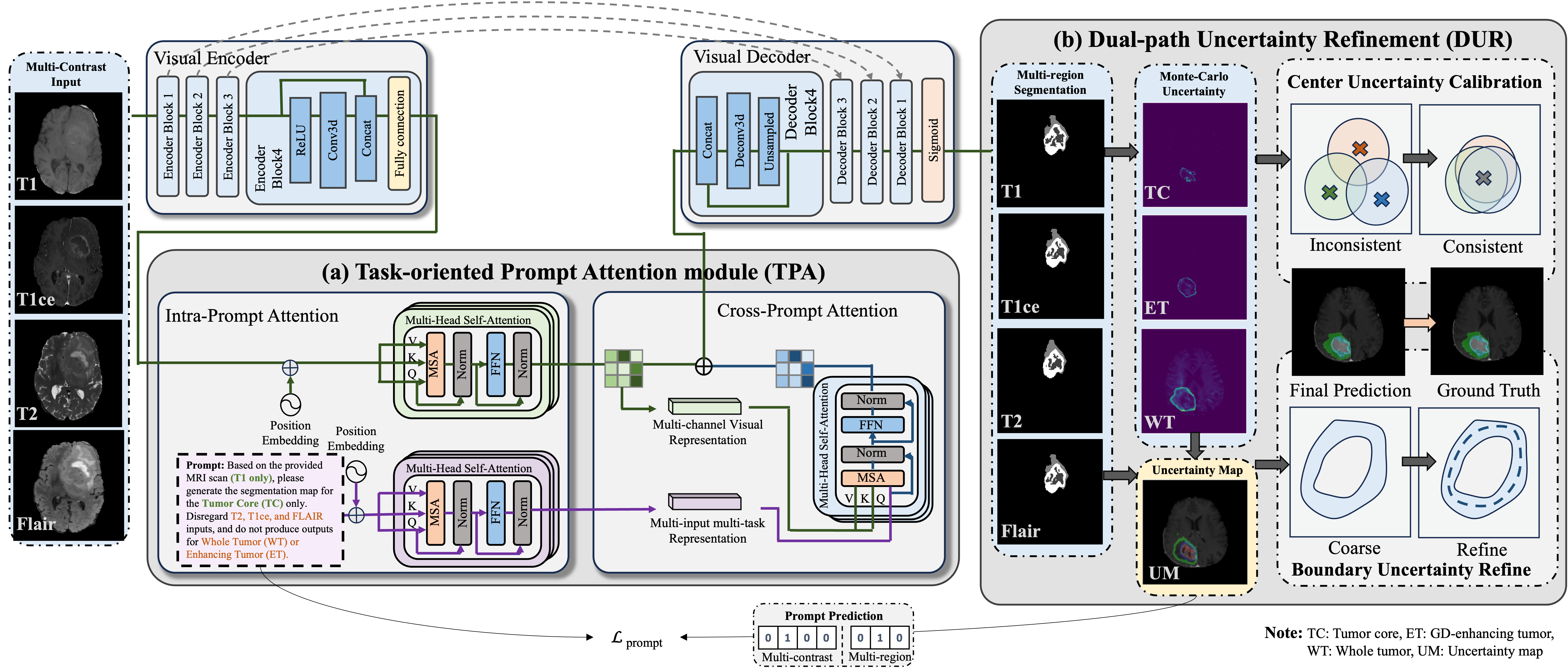}}
\caption{Workflow of the proposed Task-oriented Uncertainty Collaborative Learning (TUCL) framework. (a) The Task Prompt Attention (TPA) module integrates intra-prompt and cross-prompt attention mechanisms to capture multi-region and multi-contrast features. (b) The model leverages dual-path uncertainty refinement (DUR) to enhance segmentation accuracy. It produces consistent and refined predictions compared to ground truth. }
\label{fig2}
\end{figure*}

In this paper, we propose a Task-oriented Uncertainty Collaborative Learning (TUCL) framework (Fig. \ref{fig2}).  Compared with previous multi-input concatenation or contrast-cross attention mechanisms, we propose a task-oriented prompt and uncertainty collaborative learning mechanism. This dynamically models the contextual  relationship between multi-contrast input and multi-task output (MIMO) through the task-oriented prompt module, rather than relying solely on feature aggregation.  The cycle mapping from output prediction to text-prompt further ensures the validity of the prompt with the reverse consistency of the prediction. This makes it more robust to the case of missing modalities. Specifically, the core of our TUCL is the task-oriented prompt attention process (TPA), which consists of intra-prompt attention and cross-prompt attention. Intra-prompt attention models the multi-granularity relationships within multi-contrast and multi-region prompts independently. Cross-prompt attention fosters interaction between visual features and text prompt representations. It facilitates the collaborative learning paradigm of multi-contrast and multi-task objectives. Further, we introduce a dual-path uncertainty refinement (DUR) strategy to mitigate inconsistencies (i.e., center and boundary) and ambiguities in segmentation. It iteratively calibrates the region center and refines the boundary prediction to alleviate redundant information interference. The contribution of our work lies in three aspects:
\begin{itemize}
    \item[$\bullet$] We propose a task-oriented attention prompt and uncertainty collaborative learning mechanism to address the challenge of multi-level specificity perception in multi-contrast MRI segmentation.
    \item[$\bullet$] We design a segmentation framework achieve efficient collaboration between multi-contrast and multi-task outputs, ensuring robust performance even in the presence of missing modalities.
    \item[$\bullet$] We verify the superiority of our TUCL. It significantly improves segmentation accuracy (88.2\% in Dice and 10.853 mm in HD95) and outperforms SOTA methods.
\end{itemize}

\section{Methods}
\subsubsection*{Problem Formulation}
Let the multi-contrast MRI images be represented as a set of inputs \(\mathbf{X} = \{x_i\}_{i=1}^{N}\), where each \(x_{i} \in \mathbb{R}^{4 \times W \times H \times D}\) corresponds to a multi-contrast scan. The outputs are multi-region segmentation masks \(\mathbf{Y} = \{y_j\}_{j=1}^{M}\), where each \(y_j \in \mathbb{R}^{3 \times W \times H \times D}\) represents the segmented regions. Our objective is to learn a voxel-wise mapping function \( f_{\theta}: \mathbf{X} \rightarrow \mathbf{Y} \), parameterized by a deep neural network. The function is defined within a hypothesis space \(\mathcal{H}\), formulated as:
\begin{equation}
\mathcal{H} = \left\{ f_{\theta} \mid f: \mathbf{X} \rightarrow \mathbf{Y}, \theta \in \Theta \right\}
\end{equation}
where \(\Theta\) represents the parameter space. The optimal parameters \(\theta^*\) are obtained by minimizing a voxel-wise loss function \(\mathcal{L}\), leading to the following optimization problem:
\begin{equation}
\theta^{*}=\arg \min _{\theta \in \Theta} \mathbb{E}_{\mathbf{X} \sim \mathcal{P}_{X}, \mathbf{Y} \sim \mathcal{P}_{Y}}\left[\mathcal{L}\left(f_{\theta}(\mathbf{X}), \mathbf{Y}\right)\right]
\end{equation}
where \(\mathcal{P}_{X}\) and \(\mathcal{P}_{Y}\) denote the underlying distributions of the multi-contrast inputs and the corresponding segmentation outputs, respectively.

%\subsection{Network Implementation}
% \subsubsection{Encoder-Decoder Framework}
% The backbone of the model is a wavelet-aware encoder-decoder network that processes multi-contrast MRI inputs. The encoder extracts hierarchical features using wavelet transformation (WT) layers and convolutional layers to capture both spatial and frequency domain information. The decoder mirrors this structure with inverse wavelet deconvolution (WTDeconv3d) to reconstruct high-resolution segmentation maps. This framework ensures effective representation learning while preserving structural details.
% 
\begin{table}[t]
    \centering
    \caption{Quantitative comparison on BraTS 2021 dataset with different labeled ratios.}
    \resizebox{\columnwidth}{!}{
    \begin{tabular}{l|l|cc|cccc|cccc}
        \hline\hline
        \multicolumn{1}{l|}{\textbf{Ratio}} & \multicolumn{1}{c|}{\textbf{Method}} & \textbf{Param} & \textbf{FLOPs} & \multicolumn{4}{c}{\textbf{Dice (\%) $\uparrow$}} & \multicolumn{4}{c}{\textbf{HD95 (mm) $\downarrow$}} \\
        \cline{3-12}
        &  & (M) & (G) & ET & WT & TC & Ave & ET & WT & TC & Ave \\
        \hline
        10\%  
        & UNet  \cite{u-net} & 1.98 & 16.49 & 66.4 & 86.6 & 81.7 & 78.2 & 18.306 & 15.877 & 17.002 & 17.395 \\
        & Att-UNet  \cite{cciccek20163d} &5.91 &407.48 & 75.9 & 84.1 & 83.2 & 81.1 & 18.148 &12.626 & 19.750  & 16.175 \\
        & SegResNet\cite{hsu2021SegResNet} &4.70 &596.73 &70.1 &84.3 &82.5 & 78.9 & 20.462 &15.898 &19.676 & 18.012 \\
        & Vnet\cite{milletari2016Vnet} &45.61 &3014.13  &66.9 &80.4 &74.7 & 74.0 & 19.864 &22.050 &19.423 & 20.112 \\
        & TransBTS  \cite{wenxuan2021transbts} &32.99 &1305.92 &74.1 &78.7 &82.2 & 78.3 & 11.739 &16.796  &11.977 & 13.171 \\
        & UNETR  \cite{hatamizadeh2022unetr} &131.97 &796.87 & 75.4 & 81.4 & 75.1 & 77.3 & 17.106 &13.225  & 20.602 & 16.311 \\
        & Swin-UNETR  \cite{hatamizadeh2021swin} &62.19 &3073.72 &73.5 &85.5 &79.5 & 79.5 & 19.830 &18.152  & 21.203 & 19.728 \\
        & MedNext\cite{roy2023mednext} &6.59 &1009.47 &79.0 &87.1 &87.6 & 84.6 & 17.846 &14.471 &18.392 & 16.903 \\
        \cline{2-12}
        & Ours (w/o T1) &- &- &77.8 &88.5 &88.8 & 85.0 &15.922 &16.087 &15.175 & 15.728 \\
        & Ours (w/o T2) &- &- &77.7 &83.2 &85.1 & 82.0 &15.884 &24.039 &16.482 & 18.802 \\
        & Ours (w/o T1ce) &- &- &67.5 &88.2 &55.8 & 70.5 &18.682 &\textbf{8.425} &16.552 & 14.553 \\
        & Ours (w/o Flair) &-&-&75.8 &78.7 &84.0& 79.5 &27.654 &40.144 &29.364  & 32.387 \\
        \cline{2-12}
        & Ours &33.38 &326.8 & \textbf{80.5} & \textbf{89.1} & \textbf{88.9} & \textbf{86.2} & \textbf{13.398} & 14.524 & \textbf{12.715} & \textbf{13.546} \\
        \hline
        30\%  
        & UNet  \cite{u-net} & 1.98 & 16.49 &76.4 &89.3 &81.3 & 82.3 & 18.612 &15.209  &21.127 & 18.316 \\
        & Att-UNet  \cite{cciccek20163d} &5.91 &407.48 &77.9 &86.5 &84.1 & 82.8 & 17.733 &15.880 &20.016 & 17.876 \\
        & SegResNet\cite{hsu2021SegResNet} &4.70 &596.73 &78.0 &83.6 &80.9 & 80.8 & 17.061 &14.679 &18.355 & 16.698 \\
        & Vnet\cite{milletari2016Vnet} &45.61 &3014.13 & 76.4 &77.2 &82.9 & 78.8 & 12.901 &27.421 &14.133 & 18.818 \\
        & TransBTS  \cite{wenxuan2021transbts} &32.99 &1305.92 &76.6 &88.3 &87.5 & 84.1 & 18.328 &20.555  &17.188 & 18.357 \\
        & UNETR  \cite{hatamizadeh2022unetr} &131.97 &796.87 &73.8 &86.5 &80.9 & 80.4 & 20.700 &17.523  &22.598 & 20.607 \\
        & Swin-UNETR  \cite{hatamizadeh2021swin} &62.19 &3073.72 &75.2 &87.7 &76.7 & 79.8 & 23.475 &17.154  &27.443 & 22.357 \\
        & MedNext\cite{roy2023mednext} &6.59 &1009.47  &\textbf{82.6} &\textbf{91.4} &90.6 & 88.2 & 10.914 &\textbf{10.523}  &10.348 & 10.595 \\
        \cline{2-12}
        & Ours (w/o T1) &- &- &79.7 &73.3 &89.2 & 80.7 &12.639 &20.793 &13.474 & 15.635 \\
        & Ours (w/o T2) &- &- &82.3 &88.5 &87.0 & 85.9 &\textbf{10.342} &18.294 &11.370 & 13.335 \\
        & Ours (w/o T1ce) &- &- &54.5 &89.6 &47.4 & 63.8 &17.303 &15.111 &17.412 & 16.609 \\
        & Ours (w/o Flair) &-&-&75.2 &77.4 &84.1& 78.9 &19.835 &19.628 &19.179  & 19.547 \\
        \cline{2-12}
        & Ours &33.38 &326.8 & 81.0 & 91.3 & \textbf{92.3} & \textbf{88.2} & 10.779 & 12.202 & \textbf{9.577} & \textbf{10.853} \\
        \hline\hline
    \end{tabular}}
    \label{tab:comparison1}
\end{table}

\begin{table}[t]
    \centering
    \caption{Ablation study for different modules (TPA, DUR).}
    \resizebox{0.7\columnwidth}{!}{
    \begin{tabular}{c|ccc|cccc|cccc} % 去掉 TOC 列
        \toprule
        \textbf{Ratio} 
        & \multicolumn{3}{c|}{\textbf{Modules}} 
        & \multicolumn{4}{c|}{\textbf{Dice (\%) $\uparrow$}} 
        & \multicolumn{4}{c}{\textbf{HD95 (mm) $\downarrow$}} \\
        \cmidrule(lr){2-4} \cmidrule(lr){5-8} \cmidrule(lr){9-12}
        & Base. & TPA & DUR  
        & ET & WT & TC & Ave 
        & ET & WT & TC & Ave \\
        \midrule
        10$\%$
        & $\surd$ & - & -  
        & 63.9 & 88.3 & 83.4 & 78.5 
        & 26.190 & 14.725 & 16.429 & 19.115 \\
        & $\surd$ & $\surd$ & -  
        & 77.1 & 85.2 & 82.2 & 81.5  
        & 17.806 & 16.534 & 17.775 & 17.37 \\
        & $\surd$ & $\surd$ & $\surd$  
        & \textbf{80.5} & \textbf{89.1} & \textbf{88.9} & \textbf{86.1}  
        & \textbf{13.398} & \textbf{14.524} & \textbf{12.715} & \textbf{13.55} \\
        \midrule
        30$\%$
        & $\surd$ & - & -  
        & 78.7 & 88.4 & 83.7 & 83.6 
        & 20.126 & 12.488 & 21.963 & 18.19 \\
        & $\surd$ & $\surd$ & -  
        & \textbf{81.7} & 84.6 & 86.1 & 84.1  
        & 11.748 & 27.017 & 11.900 & 16.89 \\
        & $\surd$ & $\surd$ & $\surd$ 
        & 81.0 & \textbf{91.3} & \textbf{92.3} & \textbf{88.2} 
        & \textbf{10.779} & \textbf{12.202} & \textbf{9.577} & \textbf{10.85} \\
        \bottomrule
    \end{tabular}}
    \label{tab:comparison2}
\end{table}
\subsubsection{Task-oriented Prompt Attention Module (TPA)}
Our framework centers on a Task-oriented Prompt Attention (TPA) module that guides multi-contrast, multi-region segmentation in a task-specific manner (Fig. \ref{fig2}(a)). The TPA module consists of two components. First, intra-prompt Attention refines both the segmentation features \( f_{\mathrm{seg}} \) and the prompt features \( X_{\mathrm{prompt}} \), producing specialized representations for different contrasts and regions. Second, cross-prompt Attention fuses the refined features from both streams, enabling the prompt to steer the segmentation process and enforce consistency across modalities. This can be formally defined as:
\begin{equation}
f_{\mathrm{TPA}} = \mathrm{CrossAttn}\Bigl(\mathrm{IntraAttn}\bigl(f_{\mathrm{seg}}\bigr),\, \mathrm{IntraAttn}\bigl(X_{\mathrm{prompt}}\bigr)\Bigr).
\end{equation}
To further validate the effectiveness of the TPA module, we introduce a Task-Oriented Cycle mechanism. In this mechanism, refined segmentation predictions are fed back to update the prompt features, forming a cyclic process that iteratively enhances performance. This cycle is enforced by the following loss function:
\begin{equation}
\mathcal{L}_{\mathrm{TPA}} = \mathbb{E}_{x\in X,\, \hat{y} \in \hat{Y}} \left[\left\| X_{\mathrm{prompt}} - \Phi\bigl(F(X_{\mathrm{prompt}}), \hat{y}\bigr) \right\|^2 \right],
\end{equation}
where \( \Phi(\cdot,\cdot) \) is an update function that remaps the extracted prompt features \( F(X_{\mathrm{prompt}}) \) with the prediction \(\hat{y}\) back to the prompt space. This cyclic feedback reinforces robust prompt learning and serves as a consistency constraint, validating the TPA module’s guidance in segmentation.
% \textbf{Uncertainty Estimation and Refinement}
% The proposed method uses Monte Carlo sampling during inference to estimate pixel-wise uncertainty maps. Specifically, multiple stochastic forward passes are performed to generate probability distributions over the predictions. It can be defined as:
% $
% U=\frac{1}{T} {\textstyle \sum_{t=1}^{T}}(\hat{Y}^{(t)}-\frac{1}{T} {\textstyle\sum_{t=1}^{T}}(\hat{Y}^{(t)}))^{2}.
% $
% The resulting uncertainty map highlights regions with inconsistent predictions, which are then refined in subsequent steps.\\
% Further, the dual-path refinement (DPR) process leverages two specialized loss functions to address boundary inconsistencies and uncertainty propagation. Boundary uncertainty loss focuses on refining segmentation boundaries by penalizing discrepancies between predicted and ground truth contours in high-uncertainty regions. It ensures precise delineation of region boundaries. \\
% Uncertainty matching loss encourages the alignment of uncertainty maps with the actual segmentation errors, facilitating better calibration of the uncertainty estimates.\\
% The refinement process iteratively integrates the uncertainty maps and segmentation predictions, creating a feedback loop that improves model robustness and accuracy.
\subsubsection{Dual-path Uncertainty Refinement}
We employ dual uncertainty refinement to further refine the segmentation center and boundary. The uncertainty map uses Monte Carlo sampling to estimate pixel-by-pixel uncertainty during inference. Specifically, multiple stochastic forward passes yield a set of predictions \(\{\hat{Y}^{(t)}\}_{t=1}^T\), from which the uncertainty map \(U\) is computed as:
\begin{equation}
U = {\textstyle\frac{1}{T}\sum_{t=1}^{T}(\hat{Y}^{(t)} - \frac{1}{T}\sum_{k=1}^{T}\hat{Y}^{(k)})^{2}},
\end{equation}
where \(T\) is the number of sampling runs. The resulting uncertainty map highlights regions with inconsistent predictions. We estimate pixel-wise uncertainty via Monte Carlo sampling and partition the image domain \(\Omega\) into core and boundary regions:
$
\Omega_{c} = \{p\in\Omega \mid U(p) \leq \delta\}, \quad
\Omega_{b} = \{q\in\Omega \mid U(q) > \delta\},
$
where \(U(p)\) is the uncertainty at voxel \(p\) and \(\delta\) is a pre-defined threshold. The final loss function integrates two levels of uncertainty optimization into a unified loss.
\begin{equation}
\mathcal{L}_{\mathrm{DUR}} = {\textstyle\alpha \cdot \frac{1}{|\Omega_{c}|}\sum_{p\in\Omega_{c}} \ell\bigl(\hat{Y}(p),Y(p)\bigr)
+ \beta \cdot \frac{1}{|\Omega_{b}|}\sum_{q\in\Omega_{b}} \ell\bigl(\hat{Y}(q),Y(q)\bigr),}
\end{equation}
where \(\ell(\cdot,\cdot)\) is the overlap loss, and \(\alpha,\beta\) are weighting factors balancing the center and boundary constraints.

\begin{figure*}[t]
\centerline{\includegraphics[width=\columnwidth]{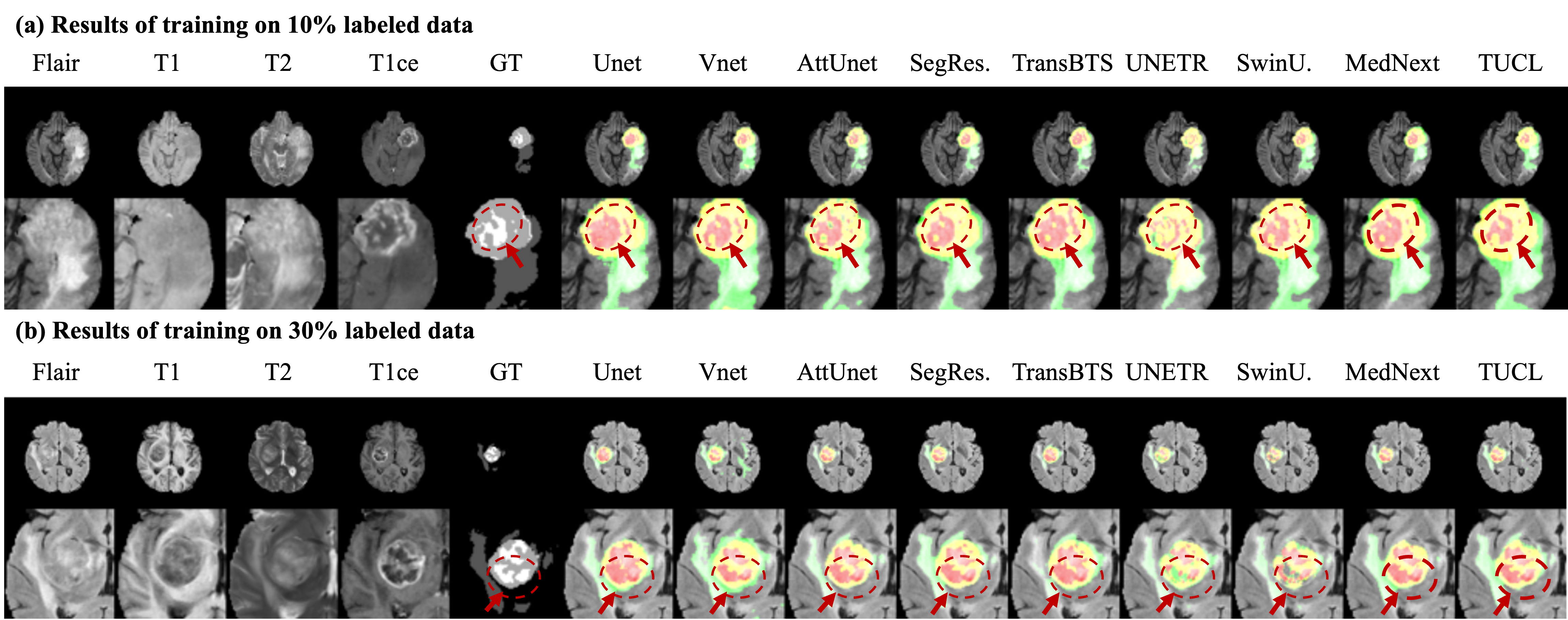}}
\caption{Comparison of brain tumor segmentation results with (a) 10\% and (b) 30\% labeled data. The first five columns show multi-contrast MRI inputs and ground truth, while the rest display segmentation results from different methods.}
\label{fig3}
\end{figure*}
\begin{figure*}[t]
\centerline{\includegraphics[width=\columnwidth]{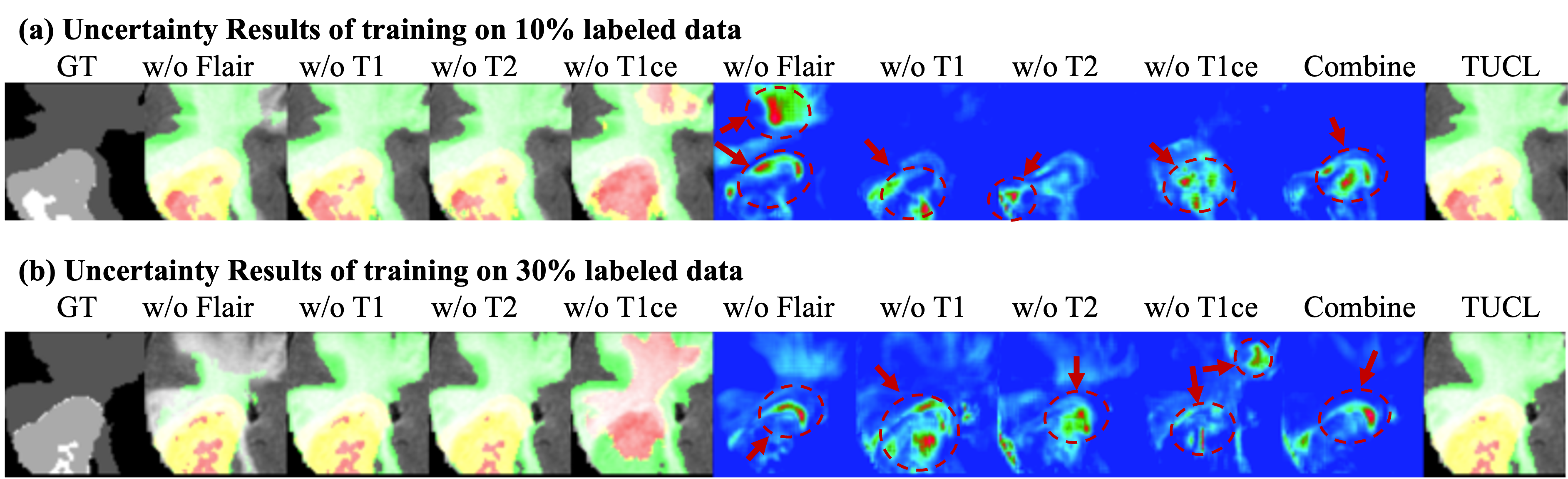}}
\caption{Uncertainty visualization of multi-contrast MRI segmentation. (a) and (b) show the results with 10\% and 30\% labeled data, respectively. The uncertainty maps highlight regions with higher prediction variability across different contrast removals.}
\label{fig:Uncertaintyvisualization4}
\end{figure*}
\begin{figure*}[t]
\centerline{\includegraphics[width=\columnwidth]{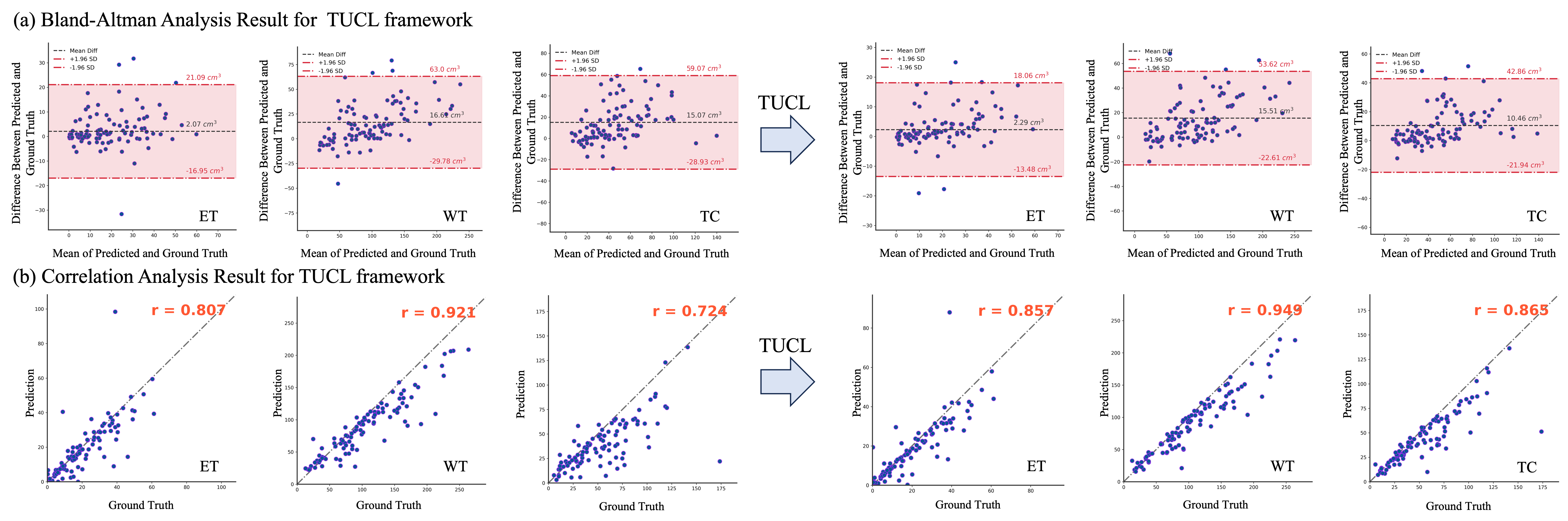}}
\caption{Bland-Altman and correlation plots of predicted versus ground truth tumor volumes. (a) The x-axis shows the mean of the predicted and true volumes, while the y-axis shows their difference; dashed lines indicate the mean bias and 95\% limits of agreement. (b) The x-axis represents the true volumes, and the y-axis represents the predicted volumes.}
\label{fig:ba5}
\end{figure*}

\subsubsection{Overall Loss Function}:
The overall loss function integrates all the components described above to achieve joint optimization of segmentation accuracy and uncertainty refinement. The total loss $\mathcal{L}_{total}$ is formulated as:
\begin{equation}
\mathcal{L}_{\text {total }}=\lambda_{1} \mathcal{L}_{\text {seg }}+\lambda_{2} \mathcal{L}_{\text {TPA }}+\lambda_{3} \mathcal{L}_{\text {DUR }},
\end{equation}
where $\lambda_{1} ,\lambda_{2} ,\lambda_{3} $ are hyperparameters determined empirically to control the contributions of each loss term.

\section{Experiment}
\textbf{Dataset:}
We use the public BRATS2021 dataset \cite{brats_2021} to evaluate the performance. The BRATS2021 dataset contains MRI data from 2040 subjects with brain tumors. The segmentation annotations comprise the GD-enhancing tumor (ET — label 4), the peritumoral edematous/invaded tissue (ED — label 2), and the necrotic tumor core (NCR — label 1). The whole tumor (WT) as the union of ED, NCR, and ET; the tumor core (TC) as the union of NCR and ET. \\
\textbf{Metric:}
We evaluate the segmentation performance of our TUCL framework using several key metrics. These include: 1) Dice Similarity Coefficient (Dice), which measures the spatial overlap between predicted and ground truth regions. 2) 95th Percentile Hausdorff Distance (HD95) to assess boundary agreement by quantifying the maximum distance between segmentation boundaries.
\\
\textbf{Comparison Experiment:}
% Table \ref{tab:comparison1} shows that our TUCL framework achieves a Dice score of 86.2\% and an HD95 of 13.546 mm with 10\% labelled data, improving to 88.2\% and 10.853 mm with 30\%. Omitting MRI modalities reduces Dice to 85.0\% (without T1), 82.0\% (without T2) and 70.5\% (without T1ce), underscoring the importance of T1ce for delineating active tumors. Removing FLAIR further drops Dice to 79.5\% and increases HD95 to 32.387 mm, highlighting the need for peritumoral oedema information. Fig. \ref{fig3} illustrates brain tumor segmentation results comparing to different other models, demonstrating that our TUCL framework consistently achieves more accurate tumor delineation (rightmost column) compared to other models.
Table \ref{tab:comparison1} and Fig. \ref{fig3} demonstrate the superior performance of our TUCL framework in brain tumor segmentation. TUCL achieves a Dice score of 86.2\% and an HD95 of 13.546 mm with 10\% labeled data,  improving to 88.2\% and 10.853 mm with 30\% labeled data. Removing MRI modalities degrades performance, with Dice dropping to 85.0\% (without T1), 82.0\% (without T2), and 70.5\% (without T1ce). These results indicate that TUCL consistently provides more accurate tumor delineation than most SOTA models, preserving fine structures and achieving competitive Dice and HD95 scores.
\\
\textbf{Ablation Study:} 
Fig. \ref{fig:Uncertaintyvisualization4} highlights increased prediction variability when specific modalities are excluded. Higher uncertainty regions, particularly in tumor boundaries, indicate the importance of T1ce and FLAIR for precise delineation. The reduction in uncertainty with 30\% labeled data suggests improved model confidence with more supervision. These findings align with Table \ref{tab:comparison2}, where incorporating the TPA and DUR modules enhances segmentation accuracy.
\\
\textbf{Clinical Metric Evaluation:}
Fig. \ref{fig:ba5} demonstrates the TUCL framework's strong predictive consistency across different tumor sub-regions. The Bland-Altman analysis (Fig. \ref{fig:ba5}a) further confirms reduced bias and narrower limits of agreement, highlighting TUCL's substantial performance enhancement in tumor volume estimation. The correlation analysis (Fig. \ref{fig:ba5}b) shows high agreement, with r-values improving from 0.724 to 0.865 (TC) and from 0.807 to 0.857(ET) after integrating TUCL.
\\
\section{Conclusion}
In this paper, we propose the Task-oriented Uncertainty Collaborative Learning (TUCL) framework to address multi-level specificity challenges in multi-contrast MRI brain tumor segmentation. TUCL integrates task-aware prompt attention and dual-path uncertainty calibration, achieving state-of-the-art performance  (Dice: 88.2\%, HD95: 10.853 mm). Although our TUCL achieves strong performance in MRI brain tumor segmentation, it still has some limitations. The TUCL needs to be extended to other diseases (e.g., Parkinson's and Alzheimer's) for broader clinical validation (e.g., CT and PET). Our future work will focus on adapting TUCL to different diseases and other imaging modalities.

\bibliographystyle{splncs04}
\bibliography{ref}

\end{document}